\documentclass[aps,twocolumn,pra,superscriptaddress,amsmath,showpacs,tightenlines]{revtex4-1}
\usepackage{amssymb}
\usepackage{amsmath}
\usepackage{graphicx}
\usepackage{subfigure}
\usepackage{natbib}
\usepackage{epsfig}
\usepackage{amsfonts}
\usepackage{mathrsfs}
\usepackage{CJK}
\usepackage{xcolor}
\usepackage[toc,page,title,titletoc,header]{appendix}
\begin{document}

\title{Bound states and  atomic interaction in giant atom waveguide QED with dispersive coupling}

\author{Mingzhu Weng}
\affiliation{Center for Quantum Sciences and School of Physics, Northeast Normal University, Changchun 130024, China}
\author{Zhihai Wang}
\email{wangzh761@nenu.edu.cn}
\affiliation{Center for Quantum Sciences and School of Physics, Northeast Normal University, Changchun 130024, China}

\begin{abstract}
In this paper, we investigate the bound states and the effective interaction between a pair of giant atoms, which couples to the coupled resonator waveguide in a nested configuration. To suppress the harmful individual and collective dissipations to the waveguide, we consider the dispersive coupling scheme, where the frequency of the giant atoms are far away from the propagating band of the waveguide. In our scheme, the atomic interaction can be induced by the overlap between the bound states in the gap. We demonstrate the relative position dependent atomic coupling and explore its application in the state transfer. We find that the transfer fidelity of a superposition state can approach $0.999$. Therefore, our scheme is useful for designing robust quantum information processing.
\end{abstract}


\maketitle
\section{introduction}
 In the rapidly evolving field of waveguide QED, the study of atom-photon interactions is of crucial importance, both in the fundamental research and the potential application fields, such as quantum computing and quantum simulations of many-body physics~\cite{DR2017,CN2017}. Different from the continuous waveguide, the coupled resonator waveguide (CRW) provides a structure for manipulating the spatial and spectral properties of photons, where the photon transport can be controlled on demand~\cite{BS2018,MZ2019,JL2019,JS2020,SP2021,KF2013}. The CRW has been realized in the platform of superconducting transmission line resonators, where the photons can be transmitted and interact with the artificial superconducting quantum bits, for example, the transmon~\cite{AB2021,IC2011,ZY2019,AS2017}.

At the scale of quantum networks~\cite{HJ2008}, the waveguides are often considered as quantum channels for photons, with the atoms (or artificial atoms) acting as quantum nodes.
When multiple atoms are coupled to a waveguide, the waveguide will serve as a data bus, to introduce the effective atomic interaction~\cite{HZ2013,XG2017}.
In the traditional waveguide QED scheme, the atom can be viewed as a point within the dipole approximation when its size is much smaller compared to the wavelength of the photons. Recently, the nonlocal coupling between the transmons and the waveguide has been demonstrated experimentally, in which the dipole approximation is no longer valid and we usually refer it as giant atom~\cite{MV2014,RM2017,AM2021}, which is a new paradigm in quantum optics.  In the giant atom scenario, the photonic interference due to the back and forth reflection between the coupling points has permitted the exploration of many amazing phenomena, such as frequency dependent atomic relaxation rates~\cite{BK2020,YT2022,XL2022}, non-exponential atomic decay~\cite{LG2017,QY2023,LD2021}, exotic atom-photon
bound states~\cite{LG2020,XW2021,KH2023,WC2022,DW2024}, non-Markovian decay dynamics~\cite{GA2019,LD2023,SG2020}, and chiral light-matter interactions~\cite{AF2018,AC2020,AS2022,XW2022,CJ2023}.

Indeed, the waveguide provides a structured environment for giant atoms.
The interaction between the giant atom and the waveguide breaks the translational symmetry of the waveguide, leading to the atom-photon bound states being located in the band gaps~\cite{WZ2020,WZ2024,XZ2023,AS2023}, where the photon is exponentially localized around the atom(s).
The previous work focused on the setting that the giant atoms as quantum nodes are located in the propagation band in the frequency domain. As a result, the dissipations play a non-negligible role and limit its application in quantum information processing although the effective inter-atom coupling can be built.

In this paper, we tackle this issue by investigating the dispersive coupling between two giant atoms and the CRW, where the frequencies of the giant atoms are set to be located in the band gap of the waveguide. Especially, we consider a nested configuration, which can not be found in the small atom counterpart. Thanks to the dispersive atom-photon coupling, we obtain the decoherence-free interaction between giant atoms, in which the individual and collective dissipations to the waveguide are perfectly suppressed. A direct application is to perform the quantum transfer between the two giant atoms, and we achieve a fidelity of $0.999$ in the transferring of coherent superposition atomic state.

\section{Theoretical Model}
\label{model}

\begin{figure}[htp]
\centering
\includegraphics[width=1\columnwidth]{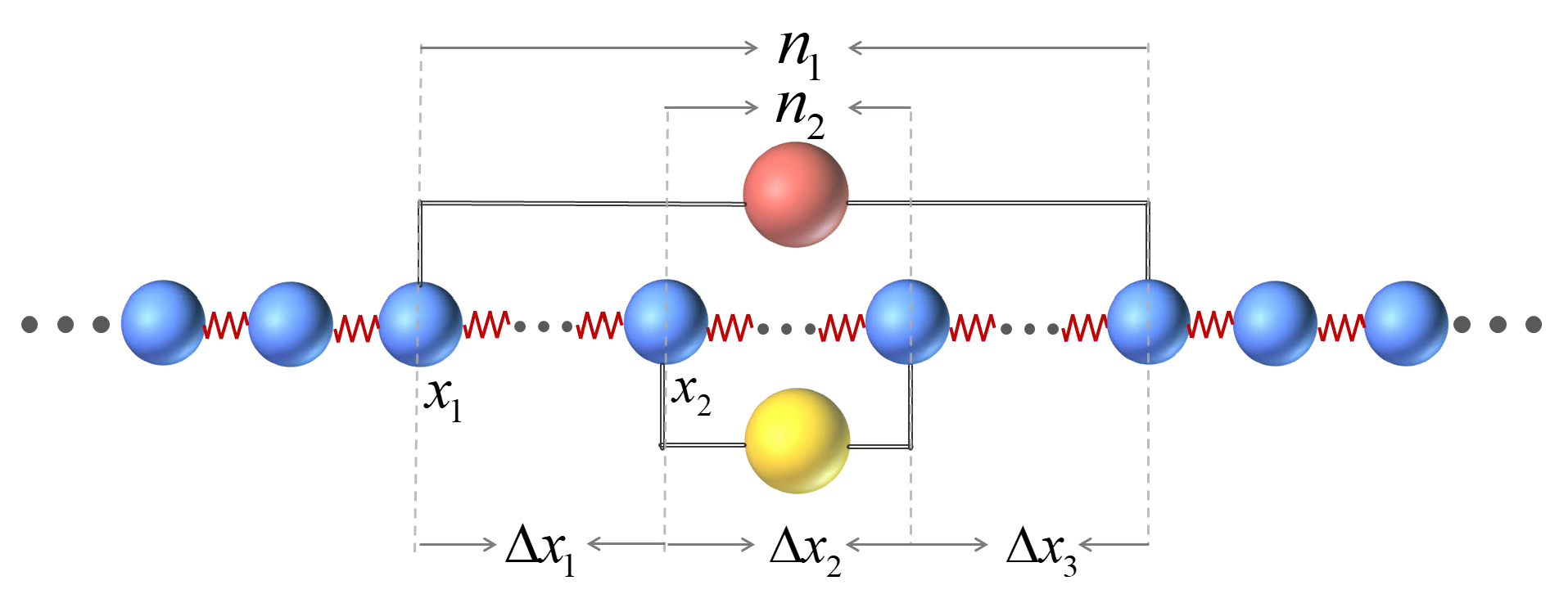}
\caption{Schematic configuration for two giant atoms that couple to a one-dimensional coupled-resonator waveguide via two sites in a nested configuration. The blue ball depicts the resonator in the waveguide and the red (yellow) ball is the giant atom labeled by $1 (2)$.}
\label{device}
\end{figure}

\begin{figure*}
\begin{centering}
\includegraphics[width=2\columnwidth]{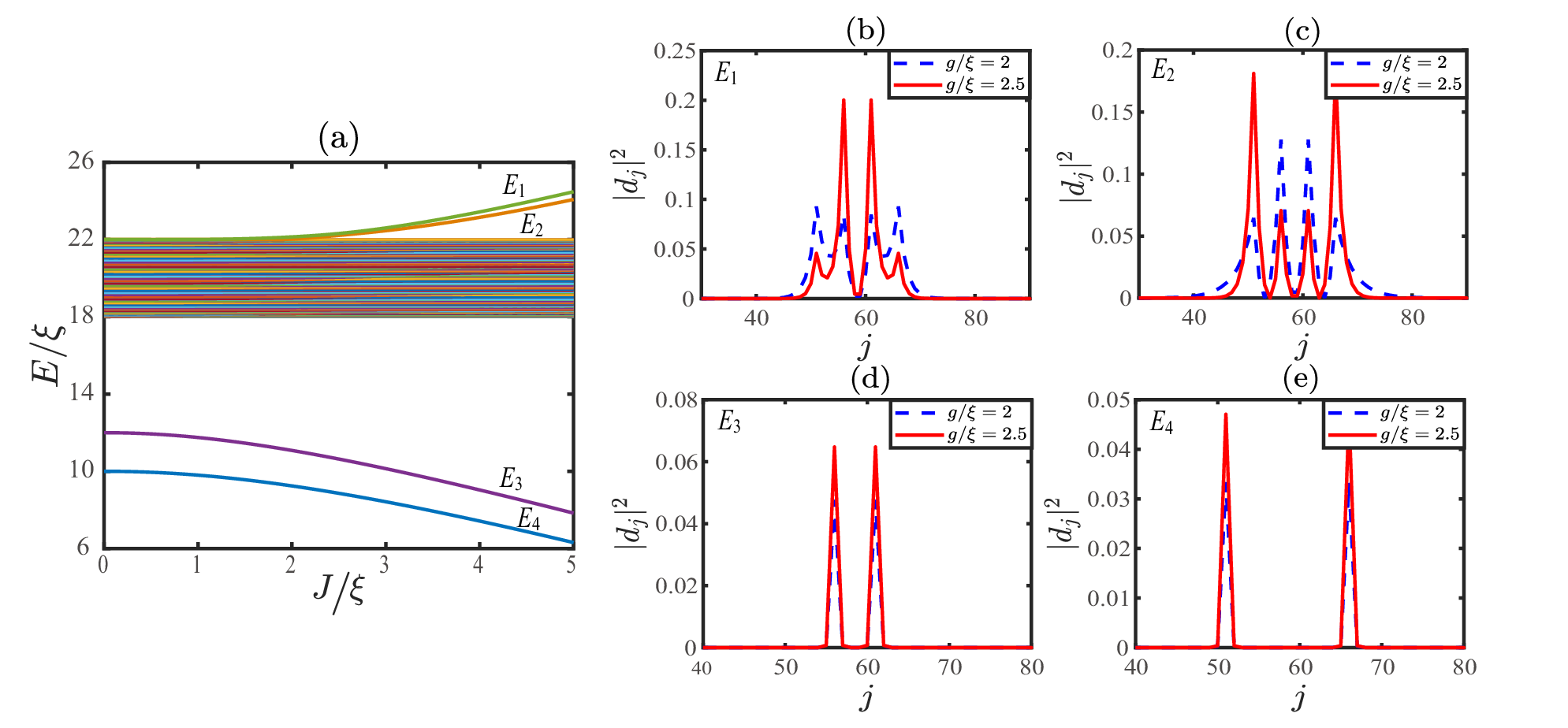}
\caption{The energy spectrum (a) and the photon population of the bound states (b-e) in two giant atoms coupled to a one dimensional waveguide with the symmetrical configuration. (b,c) corresponds to the two bound states located above the waveguide energy band. (d,e) corresponds to the two bound states located below the waveguide energy band. The parameters are set as $\omega_{c}=20\xi, \Omega_{1}=10\xi, \Omega_{2}=12\xi, n_{1}=15, n_{2}=5, \Delta x_{1}=\Delta x_{2}=\Delta x_{3}=\Delta x=5$.}
\label{energysym}
\end{centering}
\end{figure*}

 We consider a system consisting of two giant atoms which couple to a one-dimensional structured reservoir as depicted in Fig.~\ref{device}. The structured reservoir is described by the CRW with the resonators' frequency being  $\omega_c$ and the nearest-neighbor hopping strength being $\xi$. The CRW is modeled by the Hamiltonian
\begin{equation}
H_{c}=\omega_{c}\underset{j}{\sum}a_{j}^{\dagger}a_{j}
-\xi\underset{j}{\sum}(a_{j+1}^{\dagger}a_{j}+a_{j}^{\dagger}a_{j+1}).
\end{equation}
where $a_{j}^{\dagger}(a_{j})$ is the creation (annihilation) operator of the field in the waveguide on site $j$. By introducing the Fourier transformation, the Hamiltonian of the waveguide $H_c$ can be written in a diagonal form $H_c=\sum_{k}\omega_{k}a_{k}^{\dagger}a_{k}$. Here, the dispersion relation of the waveguide satisfies $\omega_{k}=\omega_{c}-2\xi\cos{k}$. Therefore, the waveguide supports a single-photon continual band which is centered at $\omega_{c}$ with the width of $4\xi$.

We further consider that two giant atoms coupled to the waveguide in a nested configuration, and the Hamiltonian is given by
\begin{eqnarray}
H_{a}&=&\Omega_{1}|e\rangle_{11}\langle e|+\Omega_{2}|e\rangle_{22}\langle e|,\\
H_{I}&=&g\sigma_{1}^{-}(a_{x_{1}}^{\dagger}+a_{x_{1}+n_{1}}^{\dagger})+g\sigma_{2}^{-}(a_{x_{2}}^{\dagger}+a_{x_{2}+n_{2}}^{\dagger})+{\rm H.c.},\nonumber
\\
\end{eqnarray}
where $H_{a}$ defines the free Hamiltonian of the two giant atoms and $H_I$ is the interaction Hamiltonian between the giant atoms and the waveguide. $\Omega_{i}$ is the transition frequency between the excited state $|e\rangle_{i}$ and the ground state $|g\rangle_{i}$ of the giant atom $i$ ($i=1,2$). The operator $\sigma_{i}^{-}(\sigma_{i}^{+})$ is the lowering (raising) operator of the giant atom $i$. $g$ is the coupling strength. By setting the distance of the nearest neighbor resonators in the waveguide as the unit of length, we can characterize the size of the giant atoms with integer index $n_{1},n_{2}$ and thus label the left (right) coupling point of $i$th giant atom as $x_{i}$ ($x_{i}+n_{i}$).

In the momentum space, the interaction Hamiltonian can be also expressed as
\begin{equation}
H_I=\frac{g}{\sqrt{N}}\sum_{i=1}^{2}\sum_k \sigma_i^{-}a_ke^{ikx_i}(1+e^{ikn_i})+{\rm H.c.}.
\end{equation}
where $N\rightarrow\infty$ is the number of resonators in the waveguide. When the frequency of the giant atoms falls into the gap of the CRW, we plot the energy spectrum by considering a symmetric configuration $\Delta x_{1}=\Delta x_{2}=\Delta x_{3}=\Delta x$ in Fig.~\ref{energysym}(a). Here, except for the continual band, we can also observe the other four curves, which are bound states located above and below the propagating band. For the upper bound states, the corresponding energy levels are separated and gradually depart from the boundary of the continual band as the atom-waveguide strength $g$ increases. The energy levels below the continual band are separated from the band even with zero atom-waveguide coupling since we have set $(\Omega_1,\Omega_2)<\omega_c-2\xi$. Furthermore, the photonic population for these four bound states $E_1\sim E_4$ are plotted in Figs.~\ref{energysym}(b-e), respectively. For the two bound states above the band, Figs.~\ref{energysym}(b,c) show the photon is bounded in the four atom-waveguide coupled sites. It is also characterized by a symmetrical distribution, that is, the height of the peak is the same for two inner and outer coupling sites, respectively. However, whether the two inner peaks or the outer peaks are higher depends on the strength of the atom-waveguide coupling. The photonic distribution for the two lower energy levels are depicted in Figs.~\ref{energysym}(d,e), which shows that the photon is only trapped in the two outer sites for one of the bound states and in the two inner two sites for the other. Both of these two states are symmetric and the photonic population increases with the coupling strength. We also investigate the bound state when the configuration of the giant atom does not satisfy the relation of $\Delta x_1=\Delta x_2=\Delta x_3$. We find that (not shown in the figure) the symmetry of photonic distribution for the two lower bound states is kept but that for the two upper ones is broken.

\section{State transfer by effective atomic coupling}

In the above section, we have demonstrated the bound states $|E_i\rangle(i=1\dots 4)$ when both of the two giant atoms couple to the CRW. Actually, if only giant atom $1$ or $2$ couples to the CRW individually, there are also bound states in which the photon is bounded near the coupled atom~\cite{WZ2020,WZ2024}. To distinguish them from the global bound states discussed in the last section, we name them as local bound states. Thus, the overlap between these local bound states will induce the indirect coupling between the giant atoms.

\begin{figure}[ht!]
\centering
\includegraphics[width=0.9\columnwidth]{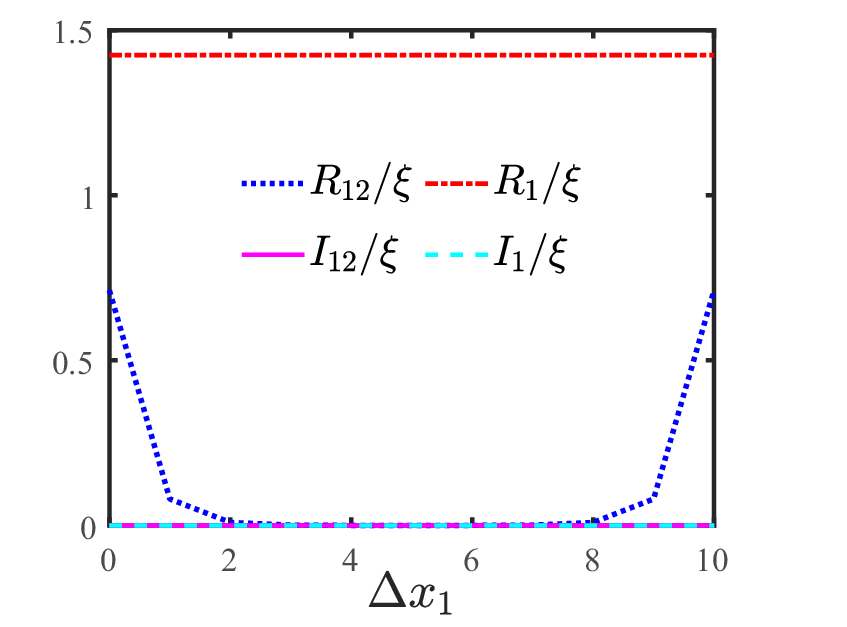}
\caption{The frequency shift $R_{1}$ and individual dissipation $I_{1}$ of the giant atom with the coupling point spacing $\Delta x_{1}$.  $R_{12}$ and $I_{12}$ are the coherent coupling and the collective dissipation between the two giant atoms, respectively. The parameters are set as $\omega_{c}=20\xi, \Omega_{1}=10\xi, \Omega_{2}=12\xi, n_{1}=15, n_{2}=5, g=2.5\xi$.}
\label{effcouple}
\end{figure}

To investigate it clearly, we write the wave function of the system in the single excitation subspace as
\begin{equation}
|\psi(t)\rangle=e^{-i\frac{\Omega_{1}+\Omega_{2}}{2}t}[\alpha_{1}(t)\sigma_{1}^{\dagger}
+\alpha_{2}(t)\sigma_{2}^{\dagger}+\sum_{k}\beta_{k}(t)a_{k}^{\dagger}]|G\rangle.
\end{equation}
where $|G\rangle$ represents that both of the two giant atoms are in the ground state while the waveguide is in the vacuum state.
Here, $\alpha_{i}(t)\,(i=1,2)$ is the amplitude for the $i$th giant atom in its excited state and $\beta_{k}(t)$ is the amplitude of a single photon state in the $k$th mode.
\begin{eqnarray}
\overset{\cdot}{\alpha_{1}}(t)&=& -i\frac{\Delta}{2}\alpha_{1}(t)-i\sum_{k}g_{1k}\beta_{k}(t),\\
\label{Alpha1}
\overset{\cdot}{\alpha_{2}}(t)&=& i\frac{\Delta}{2}\alpha_{2}(t)-i\sum_{k}g_{2k}\beta_{k}(t),\\
\label{Alpha2}
\overset{\cdot}{\beta_{k}}(t)&=& i\frac{\delta_k}{2}\beta_{k}(t)-ig_{1k}^{*}\alpha_{1}(t)-ig_{2k}^{*}\alpha_{2}(t).
\end{eqnarray}
where $\Delta=\Omega_1-\Omega_2$ is the detuning between the two giant atoms and $\delta_k$ is defined as $\delta_{k}=2\omega_{k}-(\Omega_{1}+\Omega_{2})$. $g_{ik}=g/\sqrt{N}(e^{ikx_{i}}+e^{ik(x_{i}+n_{i})})(i=1,2)$.
Since we are working in the dispersive coupling regime with $\Omega_1 \approx \Omega_2$ and $\Omega_{1(2)}-\omega_k\gg g$ for any $k\in [0,\pi)$, the parameters satisfy the relation $\delta_k\gg \Delta$.
This allows us to adiabatically eliminate the amplitudes $\beta_{k}$ by setting $\overset{\cdot}{\beta_{k}}(t)=0$.
Therefore,
\begin{equation}
\beta_{k}(t)=-\frac{2}{\delta_{k}}(g_{1k}^{*}\alpha_{1}(t)+g_{2k}^{*}\alpha_{2}(t)).
\end{equation}
Substituting back into Eqs.~(\ref{Alpha1},\ref{Alpha2}) and after some calculations, we will achieve
\begin{eqnarray}
i\dot{\alpha_{1}}=(\frac{\Delta}{2}+f_{1})\alpha_{1}+f_{12}\alpha_{2},\,\,
i\dot{\alpha_{2}}=f_{21}\alpha_{1}+(-\frac{\Delta}{2}+f_{2})\alpha_{2}.\nonumber
\\
\end{eqnarray}
where
\begin{eqnarray}
f_{i}&=&ig^{2}\int_{0}^{\infty}d\tau e^{i\delta_{c}\tau}[4J_{0}(4\xi\tau)+4i^{(n_i)}J_{n_{i}}(4\xi\tau)],\nonumber
\\
\label{fxi}
f_{12}=f_{21}&=&ig^{2}\int_{0}^{\infty}d\tau e^{i\delta_{c}\tau}[2i^{|x_{1}-x_{2}|}J_{|x_{1}-x_{2}|}(4\xi\tau)\nonumber\\
&&+2i^{|x_{1}-x_{2}-n_{2}|}J_{|x_{1}-x_{2}-n_{2}|}(4\xi\tau)\nonumber\\
&&+2i^{|x_{1}+n_{1}-x_{2}|}J_{|x_{1}+n_{1}-x_{2}|}(4\xi\tau)\nonumber\\
&&+2i^{|x_{1}+n_{1}-x_{2}-n_{2}|}J_{|x_{1}+n_{1}-x_{2}-n_{2}|}(4\xi\tau)],
\label{fx1x2}
\end{eqnarray}
with $\delta_{c}=2\omega_{c}-(\Omega_{1}+\Omega_{2})$. $f_i$ only depends on the size of the $i$th giant atom $n_{i}$. The real and imaginary parts represent the frequency shift and dissipation, respectively. The results in Fig.~\ref{effcouple} demonstrate the finite frequency shift and nearly zero dissipation. Moreover, the real part  ($R_{12}={\rm Re}(f_{12})$) and imaginary part ($I_{12}={\rm Im}(f_{12})$) represents the coherent coupling and collective dissipation between the two separated giant atoms, which is induced by the spatial overlap of the local bound states. In Fig.~\ref{effcouple}, we plot $R_{12}$ as a function of their relative position, which is characterized by $\Delta x_1$, by fixing the size of both of the giant atoms $n_1$ and $n_2$. It shows that the effective coupling strength first decreases and then gradually increases with the movement of the inner giant atom. When the inner giant atom meets the coupling site of the outer giant atom ($\Delta x_{1}=0$ and $\Delta x_{1}=10$), the coupling strength is significantly enhanced and reaches the maximum. Meanwhile, the result for $I_{12}$ shows that the collective dissipation is always suppressed. Actually, such a decoherence free interaction between giant atoms can also be achieved when their frequencies fall into the continual band of the waveguide~\cite{AC2020,AF2018,MW2024}, but it requires strict parameter conditions. In our dispersive atom-waveguide coupling scheme, the simultaneous suppression of the individual and collective dissipation is robust to the geometric configuration, and therefore much more beneficial for quantum information processing.

\begin{figure}
\begin{centering}
\includegraphics[width=1\columnwidth]{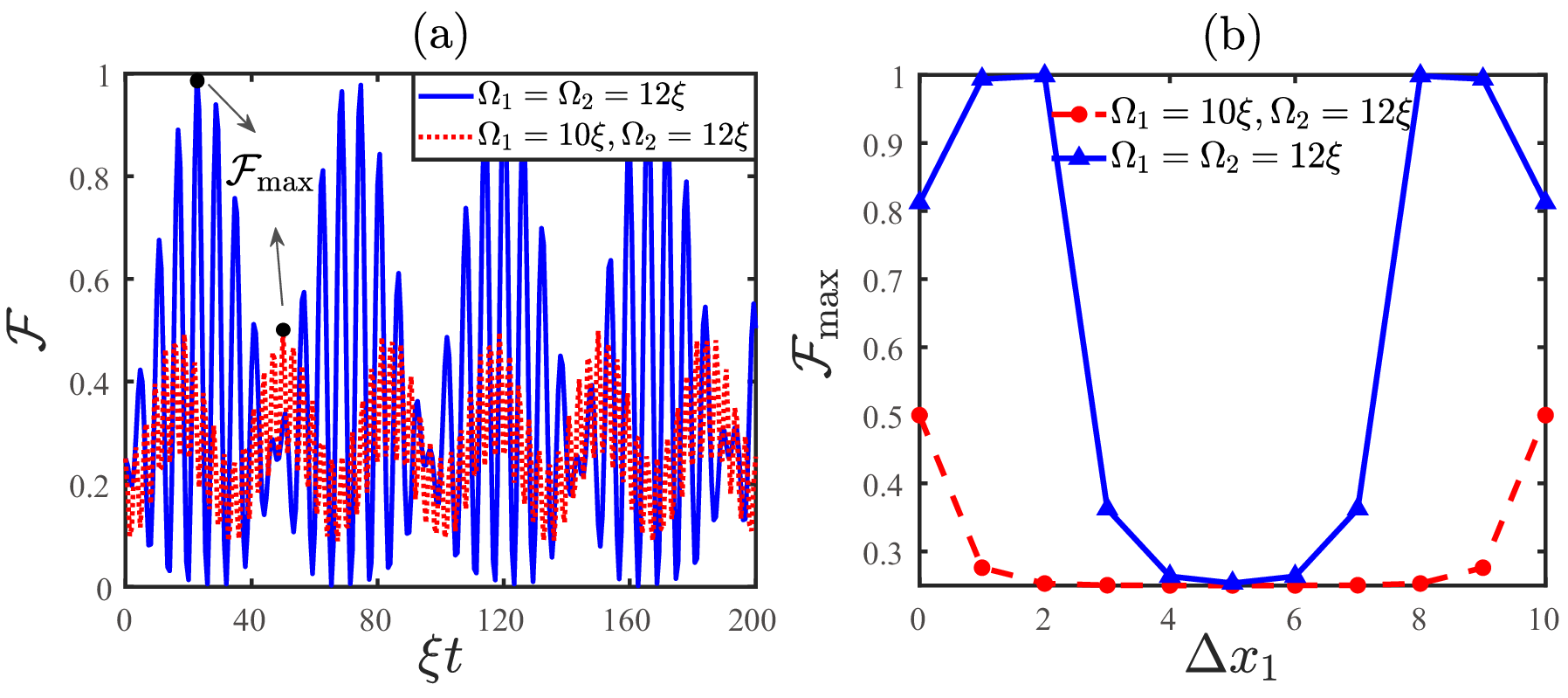}
\caption{(a) Time evolution of the fidelity $\mathcal{F}$, where the black dot denotes the maximum value $\mathcal{F}_{\rm max}$ of the fidelity $\mathcal{F}$. (b) The maximal fidelity $\mathcal{F}_{\rm max}$ versus the coupling point spacing $\Delta x_{1}$. The other parameters are set as $\omega_{c}=20\xi, n_{1}=15, n_{2}=5, g=2\xi$.}
\label{fidelity}
\end{centering}
\end{figure}

A typical example is the state transfer between the two giant atoms. We initially set the system at the state $|\psi_{0}\rangle=(|g\rangle_{1}+|e\rangle_{1})\otimes|g\rangle_{2}/\sqrt{2}$ and aim to transfer the superposition character of the first giant atom to the second one. In Fig.~\ref{fidelity}(a), we plot the time evolution of the fidelity $\mathcal{F}=|\langle\psi (t)|\psi_{T}\rangle|^{2}$ with the target state being $|\psi_{T}\rangle=|g\rangle_{1}\otimes(|g\rangle_{2}+|e\rangle_{2})/\sqrt{2}$, and it is characterized by the periodical behavior.  To visualize the effect of the atomic configuration on the state transfer, we further plot the maximal fidelity $\mathcal{F}_{\rm max}$ [labeled by the black point in Fig.~\ref{fidelity}(a)] versus the coupling point spacing $\Delta x_{1}$ in Fig.~\ref{fidelity}(b). The red dashed line in Fig.~\ref{fidelity}(b) is consistent with the blue dashed line in Fig.~\ref{effcouple},  implying that the atomic coupling leads to the quantum state transfer. In Fig.~\ref{fidelity}, the blue curve shows that the state transfer process can be further optimized to enlarge the fidelity as high as $\mathcal{F}_{\rm max}=0.9986$ when the two giant atoms are resonant in frequency for certain spacing $\Delta x_1$. For other spacing, the fidelity can not be enhanced obviously even in the resonant case due to the weak effective atomic coupling.

\section{Conclusion}

The giant atom has been experimentally realized by coupling superconducting transmon quantum qubit to the surface acoustic wave~\cite{MV2014,RM2017} or bent transmission line waveguide~\cite{BK2020,GA2019,ZW2022}. Meanwhile, with the platform of superconducting materials, CRW has been realized by the high-impedance microwave resonators and the nearest hopping strength has been achieved from $50$ MHz to $200$ MHz~\cite{PR2017,XYZ2023}.  In such systems, the qubit-waveguide coupling strength can be achieved by approximately $300$ MHz~\cite{MS2022}.

In conclusion, we have investigated the bound states when two giant atoms are coupled to a CRW in the nested configuration. We consider that both of the giant atoms are largely detuned from the continual band of the CRW, that is, the giant atoms dispersively couple to the  CRW. As a result, the virtual photon exchange with the CRW leads to the effective coupling between the two giant atoms. We interestingly find that both the individual and collective dissipation are significantly suppressed due to the dispersive coupling. Compared to the case when the giant atoms are located in the CRW continual band in frequency, our scheme releases the strict parameter condition for constructing such decoherence-free interaction. A direct application is the quantum state transfer from one atom to the other and the fidelity can achieve approximately $0.999$. Therefore, the bound state induced interaction between giant atoms in the dispersive regime can be used in quantum information processing.

\begin{acknowledgments}
Z.W. is supported by the Science and Technology Development Project of Jilin
Province (Grants No. 20230101357JC and 20220502002GH), National Science Foundation of China (Grant No. 12375010), and the Innovation Program for Quantum Science and Technology (No. 2023ZD0300700).
\end{acknowledgments}


\begin{thebibliography}{99}
\bibitem{DR2017}D. Roy, C. M. Wilson, and O. Firstenberg, Rev. Mod. Phys. {\bf 89}, 021001 (2017).

\bibitem{CN2017}C. Noh and D. G. Angelakis, Rep. Prog. Phys. {\bf 80}, 016401 (2017).

\bibitem{BS2018}B. Stern, X. Ji, Y. Okawachi, A. L. Gaeta, and M. Lipson, Nature {\bf 562}, 401 (2018).

\bibitem{MZ2019}M. Zhang, B. Buscaino, C. Wang, A. Shams-Ansari, C. Reimer, R. Zhu, J. M. Kahn, and
M. Lon\~{c}ar, Nature {\bf 568}, 373 (2019).

\bibitem{JL2019}J. Lu, J. B. Surya, X. Liu, A. W. Bruch, Z. Gong, Y. Xu, and H. X. Tang, Optica {\bf 6}, 1455
(2019).

\bibitem{JS2020}J. Szabados, D. N. Puzyrev, Y. Minet, L. Reis, K. Buse, A. Villois, D. V. Skryabin, and
I. Breunig, Phys. Rev. Lett. {\bf 124}, 203902 (2020).

\bibitem{SP2021}S.-P. Yu, D. C. Cole, H. Jung, G. T. Moille, K. Srinivasan, and S. B. Papp, Nature Photonics
{\bf 15}, 461 (2021).

\bibitem{KF2013}K. Fang and S. Fan, Phys. Rev. Lett. {\bf 111}, 203901 (2013).

\bibitem{AB2021}A. Blais, A. L. Grimsmo, S. M. Girvin, and A. Wallraff, Rev. Mod. Phys. {\bf 93}, 025005 (2021).

\bibitem{IC2011}I.-C. Hoi, C. M. Wilson, G. Johansson, T. Palomaki, B. Peropadre, and P. Delsing, Phys. Rev.
Lett. {\bf 107}, 073601 (2011).

\bibitem{ZY2019}Z. Yan, Y.-R. Zhang, M. Gong, Y. Wu, Y. Zheng, S. Li, C. Wang, F. Liang, J. Lin, Y. Xu,
C. Guo, L. Sun, C.-Z. Peng, K. Xia, H. Deng, H. Rong, J. Q. You, F. Nori, H. Fan, X. Zhu,
and J.-W. Pan, Science {\bf 364}, 753 (2019).

\bibitem{AS2017}A. Stockklauser, P. Scarlino, J. V. Koski, S. Gasparinetti, C. K. Andersen, C. Reichl,
W. Wegscheider, T. Ihn, K. Ensslin, and A. Wallraff, Phys. Rev. X {\bf 7}, 011030 (2017).

\bibitem{HJ2008}H. J. Kimble, Nature (London) {\bf 453}, 1023 (2008).

\bibitem{HZ2013}H. Zheng and H. U. Baranger, Phys. Rev. Lett. {\bf 110}, 113601 (2013).

\bibitem{XG2017}X. Gu, A. F. Kockum, A. Miranowicz, Y.-X. Liu, and F. Nori, Phys. Rep. {\bf 718-719}, 1 (2017).

\bibitem{MV2014}M. V. Gustafsson, T. Aref, A. F. Kockum, M. K. Ekstr{\"o}m, G.
Johansson, and P. Delsing, Science {\bf 346}, 207 (2014).

\bibitem{RM2017}R. Manenti, A. F. Kockum, A. Patterson, et al., Nat. communications {\bf 8}, 975 (2017).

\bibitem{AM2021}A. M. Vadiraj, A. Ask, T. G. McConkey, I. Nsanzineza, C. W.
Sandbo Chang, A. F. Kockum, and C. M. Wilson, Phys. Rev. A {\bf 103}, 023710 (2021).


\bibitem{BK2020}B. Kannan, M. Ruckriegel, D. Campbell, A. F. Kockum, J. Braum{\"u}ller, D. Kim, M. Kjaergaard, P. Krantz, A. Melville, B. M. Niedzielski, A. Veps{\"a}l{\"a}inen, R. Winik, J. Yoder, F. Nori, T. P. Orlando, S. Gustavsson, and W. D. Oliver, Nature (London) {\bf 583}, 775 (2020).

\bibitem{YT2022}Y. T. Chen, L. Du, L. Guo, Z. Wang, Y. Zhang, Y. Li, and J. H. Wu, Commun. Phys. {\bf 5}, 215 (2022).

\bibitem{XL2022}X.-L. Yin, Y.-H. Liu, J.-F. Huang, and J.-Q. Liao,  Phys. Rev. A {\bf 106}, 013715 (2022).


\bibitem{LG2017}L. Guo, A. L. Grimsmo, A. F. Kockum, M. Pletyukhov, and G. Johansson, Phys. Rev. A {\bf 95}, 053821 (2017).

\bibitem{LD2021}L. Du, M. Cai, J. Wu, Z. Wang, and Y. Li, Phys. Rev. A {\bf 103}, 053701 (2021).

\bibitem{QY2023}Q.-Y. Qiu, Y. Wu, and X.-Y. Lu, Sci. China Phys. Mech. Astron. {\bf 66}, 224212 (2023).

\bibitem{LG2020}L. Guo, A. F. Kockum, F. Marquardt, and G. Johansson, Phys. Rev. Research {\bf 2}, 043014 (2020).


\bibitem{XW2021}X. Wang, T. Liu, A. F. Kockum, H.-R. Li, and F. Nori, Phys. Rev. Lett. {\bf 126}, 043602 (2021).

\bibitem{KH2023}K. H. Lim, W.-K. Mok, and L.-C. Kwek, Phys. Rev. A {\bf 107}, 023716 (2023).

\bibitem{WC2022}W. Cheng, Z. Wang, and Y.-x. Liu, Phys. Rev. A {\bf 106}, 033522 (2022).

\bibitem{DW2024}D. W. Wang, C. Zhao, Y. T. Yan, J. Yang, Z. Wang, and L. Zhou, Phys. Rev. A {\bf 109}, 053720 (2024).

\bibitem{GA2019}G. Andersson, B. Suri, L. Guo, T. Aref, and P. Delsing, Nat. Phys. 15, 1123 (2019).

\bibitem{LD2023}L. Du, L. Guo, and Y. Li, Phys. Rev. A {\bf 107}, 023705 (2023).

\bibitem{SG2020}S. Guo, Y. Wang, T. Purdy, and J. Taylor, Phys. Rev. A {\bf 102}, 033706 (2020).

\bibitem{AF2018}A. F. Kockum, G. Johansson, and F. Nori, Phys. Rev. Lett. {\bf 120}, 140404 (2018).

\bibitem{AC2020}A. Carollo, D. Cilluffo, and F. Ciccarello, Phys. Rev. Research {\bf 2}, 043184 (2020).

\bibitem{AS2022}A. Soro, and A. F. Kockum, Phys. Rev. A {\bf 105}, 023712 (2022).

\bibitem{XW2022}X. Wang and H.-R. Li, Quantum Sci. Technol. {\bf 7}, 035007 (2022).

\bibitem{CJ2023}C. Joshi, F. Yang, and M. Mirhosseini, Phys. Rev. X {\bf 13}, 021039 (2023).

\bibitem{WZ2020}W. Zhao and Z. Wang, Phys. Rev. A {\bf 101}, 053855 (2020).

\bibitem{WZ2024}W. Z. Jia and M. T. Yu, Opt. Express {\bf 32}, 9495 (2024).

\bibitem{XZ2023}X. Zhang, C. Liu, Z. Gong, and Z. Wang, Phys. Rev. A {\bf 108}, 013704 (2023).

\bibitem{AS2023} A. Soro, C. S. Mu{\~n}oz, and A. F. Kockum, Phys. Rev. A {\bf 107}, 013710 (2023).

\bibitem{MW2024}M. Weng, X. Wang, and Z. Wang, Phys. Rev. A {\bf 110}, 023721 (2024).

\bibitem{ZW2022}Z.-Q. Wang, Y.-P. Wang, J. Yao, et al., Nat. communications {\bf 13}, 7580 (2022).

\bibitem{PR2017}P. Roushan, C. Neill, J. Tangpanitanon, et al., Science {\bf 358}, 1175 (2017).

\bibitem{XYZ2023}X. Zhang, E. Kim, D. K. Mark, S. Choi, and O. Painter, Science {\bf 379}, 278 (2023).

\bibitem{MS2022}M. Scigliuzzo, G. Calaj\'{o}, F. Ciccarello, D. P. Lozano, A. Bengtsson, P. Scarlino, A. Wallraff, D. Chang, P. Delsing, and S. Gasparinetti, Phys. Rev. X {\bf 12}, 031036 (2022).










\end{thebibliography}
\end{document}